# Polarized Superradiance from CsPbBr$_3$ Quantum Dot Superlattice with Controlled Inter-dot Electronic Coupling

*Lanyin Luo[1,2], Xueting Tang[3], Junhee Park[3], Chih-Wei Wang[3], Mansoo Park[4], Mohit Khurana[1,2], Ashutosh Singh[1], Jinwoo Cheon[4], Alexey Belyanin[1], Alexei V. Sokolov[1,2] and Dong Hee Son[1,3,4,*]*

[1]Department of Physics and Astronomy, Texas A&M University, College Station, TX 77843, USA.
[2]Institute of Quantum Science and Engineering, Texas A&M University, College Station, TX 77843, USA.
[3]Department of Chemistry, Texas A&M University, College Station, Texas 77843, USA.
[4]Center for Nanomedicine, Institute for Basic Science and Graduate Program of Nano Biomedical Engineering, Advanced Science Institute, Yonsei University, Seoul 03722, Republic of Korea.

[*]Corresponding author E-mail: dhson@tamu.edu

**Cooperative emission of photons from an ensemble of quantum dots (QDs) as superradiance can arise from the electronically coupled QDs with a coherent emitting excited state. This contrasts with superfluorescence (Dicke superradiance), where the cooperative photon emission occurs via a spontaneous buildup of coherence in an ensemble of incoherently excited QDs via their coupling to a common radiation mode. While superfluorescence has been observed in perovskite QD systems, reports of superradiance from the electronically coupled ensemble of perovskite QDs are rare. Here, we demonstrate the generation of polarized superradiance with a very narrow linewidth (<5 meV) and a large redshift (~200 meV) from the electronically coupled CsPbBr$_3$ QD superlattice achieved through a combination of strong quantum confinement and ligand engineering. In addition to photon bunching at low excitation densities, the superradiance is polarized in contrast to the uncoupled exciton emission from the same superlattice. This finding suggests the potential for obtaining polarized cooperative photon emission via anisotropic electronic coupling in QD superlattices even when the intrinsic anisotropy of exciton transition in individual QDs is weak.**

The phenomenon of cooperative emission from an ensemble of dipoles arises from the creation a coherent macroscopic dipole that produces a short burst of spontaneous emission first introduced by Dicke.[1] In the case of Dicke superradiance, sometimes called superfluorescence, the correlation between optical dipole oscillations of individual emitters is established via their interaction with a common radiation field after the excitation of an incoherent ensemble of absorbers. Therefore, superfluorescence occurs at excitation intensities that can prepare a sufficiently large number of emitters, exhibiting threshold behavior, with a finite delay time required to establish phase coherence among the emitters.[2-5] However, cooperative emission of photons may also arise from direct electronic coupling between individual emitters, as in the case of J-aggregates in molecular systems.[6-9] Superradiance from the electronically coupled coherent emitting state can develop without delay, does not exhibit threshold behavior, and should appear at significantly lower excitation intensities than those required for superfluorescence. Furthermore, in superfluorescence (Dicke superradiance), the radiative decay rate can be enhanced by many orders of magnitude up to a factor of $N$, where $N$ is the number of quantum emitters that are synchronized.[2,3] In contrast, in superradiance from electronically coupled emitters, the optical dipole matrix element between the resulting electron bands is determined by the overlap of electron orbitals or the hopping parameter in the tight-binding picture.[10,11] At the same time, photon bunching is expected as a universal signature of the cooperative emission in any coupling scenario.[2-4,12] Therefore, electronically coupled systems offer greater flexibility and control over the structure of the emitting states and the properties of the cooperative emission, which is easier to achieve and more robust with respect to decoherence under weak excitation conditions.

Metal halide perovskite nanocrystals possess beneficial features as a source of photons such as high luminescence quantum yield and facile tunability of bandgap.[13,14] Superfluorescence has been observed from various metal halide perovskite nanocrystals and 2D sheets in recent years.[12,15,16] More recently, superradiance from the electronically coupled ensemble of perovskite nanocrystals has been reported in the superlattice of $CsPbBr_3$ quantum dots (QDs), exhibiting varying degrees of coherence depending on disorder and defects within the superlattice.[17] Since the overlap of the exciton wavefunctions is crucial for creating a coherently coupled excited state, both the spatial proximity between the QDs and the quantum confinement of the QDs forming superlattice should play an important role in producing superradiance. So far, cooperative photon emission from $CsPbBr_3$ QDs were observed in 3D superlattices of weakly quantum-confined QDs passivated with relatively long surface ligands that limit the extent of spatial overlap of the exciton wavefunctions.[12,17] Here, we report the polarized superradiance exhibiting very narrow linewidth (<5 meV) and large spectral redshift (~200 meV) achieved via a combination of the strong quantum confinement imposed on the highly uniform ensemble of QDs and ligand engineering on the QD surface. This strategy not only enhanced the electronic coupling but also introduced unexpected anisotropy, despite the absence of strong intrinsic asymmetry in the QDs, enabling polarized superradiance that is more suitable for applications in photonic devices than randomly polarized emission.

To investigate how the control of quantum confinement and spatial overlap of exciton wavefunctions alter the superradiant properties of a QD superlattice, $CsPbBr_3$ QDs of two different sizes passivated with ligands of two different lengths were prepared. 9 nm and 4 nm $CsPbBr_3$ QDs with highly uniform size and shape, which are in weak and strong quantum confinement regime respectively (exciton Bohr radius: ~3.5 nm), were synthesized as described in the Method section. Oleylammonium bromide (OLAB) with 18 carbons and a bidentate ligand with 8 carbons (3C-C8) were used to vary the facet-to-face distance between the QDs in the superlattice. Fig. 1a-1d compare the solution-phase absorption and photoluminescence (PL) spectra of the QDs at room temperature that show the quantum confinement effect on exciton transition energy. Representative transmission electron microscope (TEM) images of the QDs and the superlattice formed from each QD are shown in Fig. 1e-1l. QDs passivated with OLAB and 3C-C8 provide facet-to-facet distances of 3-2.5 nm and ~1.4 nm respectively in the close-packed assembly based on TEM image analysis. Details of superlattice preparation and structural characterization are provided in Supplementary Information (SI), see Fig. S1-S3.

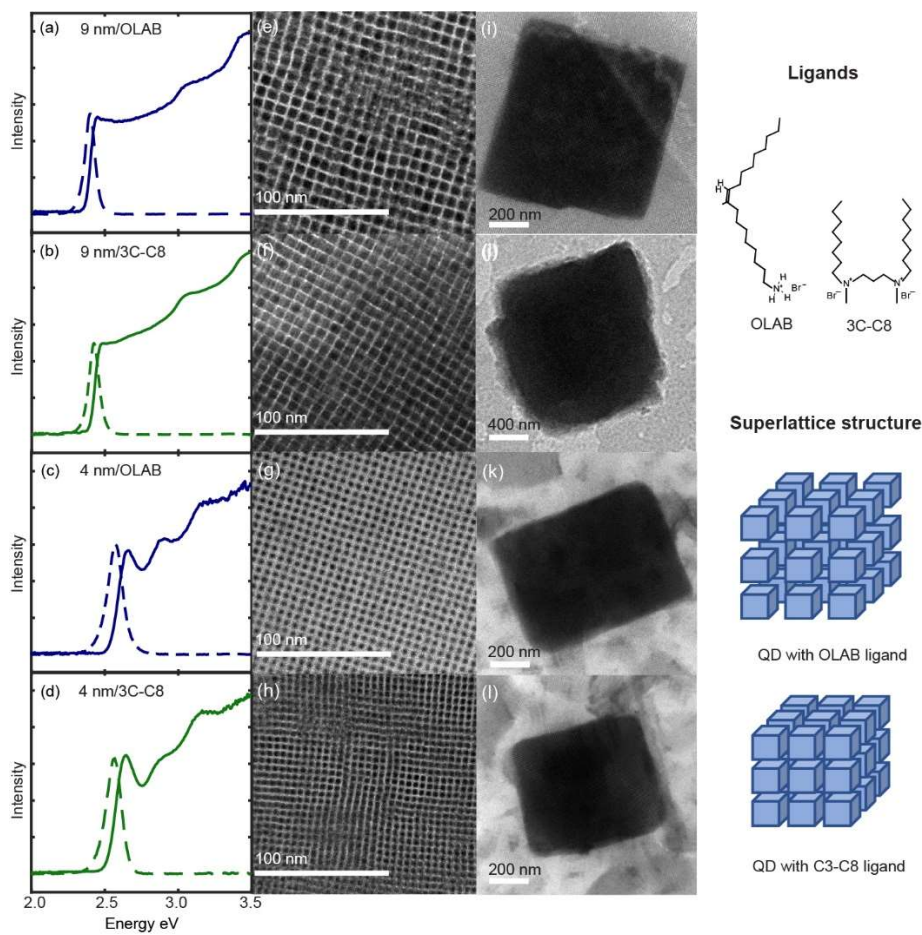

**Fig. 1** (a-d) Solution-phase absorption and photoluminescence spectra of $CsPbBr_3$ QDs of different sizes and passivating ligands. Each panel is labeled with QD size/ligand, (a) 9 nm/OLAB, (b) 9 nm/3C-C8, (c) 4 nm/OLAB, and (d) 4 nm/3C-C8. (e-h) TEM images of QD sample on TEM grid, (i-l) TEM image of a superlattice formed from each QD sample. The right side of the figure shows the chemical structure of OLAB and 3C-C8 ligands and the illustration of the superlattice structures with ligand-tuned facet-to-facet distance between the QDs.

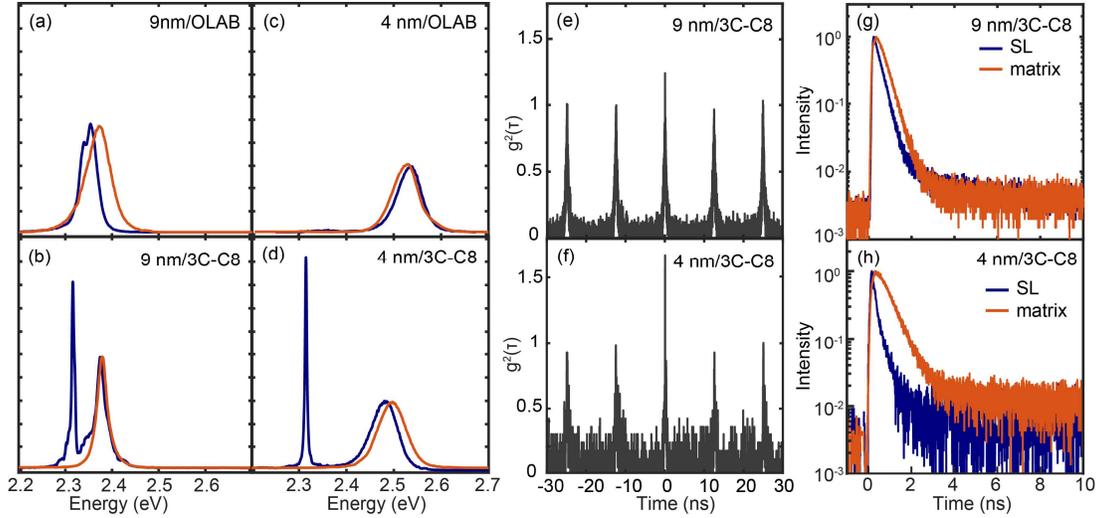

**Fig. 2** (a-d) PL spectra of superlattice (blue) and dilute dispersion (red) of QDs measured at 10 K. The label inside each panel represents the QD size/ligand. (e, f) Second-order photon correlation, $g^2(\tau)$, of superradiance from the superlattice formed from (e) 9 nm/3C-C8 and (f) 4 nm/3C-C8 QDs. (g, h) Comparison of the normalized time-dependent PL intensities of superradiance from the superlattice and diluted dispersion of QDs in polymer matrix, (g) 9 nm/3C-C8 and (h) 4 nm/3C-C8 QDs.

The blue curves in Fig. 2a-2d are the PL spectra at 10 K measured from superlattices formed from the four different QDs shown in Fig. 1. For comparison, the PL spectra from a dilute dispersion of the same QDs in polystyrene matrix are shown in red. 405 nm excitation at the fluence of 240 nJ/cm$^2$ per pulse and the repetition rate of 5 MHz was used to ensure sufficiently low excitation density (0.03 and 0.003 per uncoupled QD for 9 nm and 4 nm QDs respectively based on the reported absorption cross section[18]). With OLAB ligand, 4 nm QDs show no noticeable sign of coupling between the QDs in their PL spectra, while 9 nm QDs show a small redshift indicative of some electronic coupling similar to what has been observed in a earlier study.[17] With 3C-C8 ligands, on the other hand, both 4 nm and 9 nm QDs display an additional redshifted peak with narrower linewidth. The reversible appearance and disappearance of this new peak upon temperature cycling, along with a clear difference from the bulk-like CsPbBr$_3$ nanocrystals in the temperature-dependent peak position and linewidth,[19, 20] rule out the merging of QDs in the superlattice. These new peaks are attributed to the superradiance from the coupled QDs in the superlattice. The higher-energy peaks are similar to those from the QDs dispersed in polystyrene matrix, which are attributed to PL from a subpopulation of QDs not coupled in the superlattice.[12, 21] The 4 nm/3C-C8 QD superlattices typically exhibited the full width at half maximum (FWHM) linewidth of 3-5 meV and the redshift of 180-220 meV relative to the uncoupled exciton PL (See Fig. S4 in SI). These are the narrowest linewidth and the largest redshift observed to date for cooperative photon emission from CsPbBr$_3$ QDs as either superfluorescence or superradiance. The 9 nm/3C-C8 QD superlattice exhibited 7-10 meV linewidth and ~70 meV redshift. While the linewidth narrowing and the redshift of the superradiance is less pronounced than in 4 nm/3C-C8 QD superlattice, these are comparable to those of superfluorescence recently reported for the superlattice made with OLAB-passivated QDs of similar size.[12, 17] If the linewidth and redshift are taken as the indicators of the extent of coupling between the QDs, the contrast in the PL spectra between OLAB- and 3C-C8-passivated QDs highlights the importance of ligand tuning in obtaining the QD coupling necessary to produce superradiance.

The features of cooperative emission from coupled QDs include photon bunching and an accelerated radiative decay rate. Photon bunching was confirmed through the second-order photon correlation, $g^2(\tau)$, measurements using a Hanbury Brown-Twiss interferometer. Spontaneous emission from uncorrelated emitters shows the random Poisson distribution of photon arrival times, giving an average $g^2(\tau)$ value of 1, while photons from correlated emitters tend to bunch with $g^2(\tau) < g^2(0)$ and $g^2(0) > 1$.[22, 23] Fig. 2e and 2f show the $g^2(\tau)$ profiles of superradiance from the superlattices formed from 9 nm/3C-C8 and 4 nm/3C-C8 QDs measured at 10 K, both of which exhibit photon bunching at τ = 0. For these measurements, a 405 nm picosecond pulsed laser at the repetition rate of 80 MHz and photon fluence of 240 nJ/cm$^2$ was used. 4 nm/3C-C8 QDs show $g^2(0) = 1.6$, significantly larger than that of 9 nm/3C-C8 QDs. This indicates the higher cooperativity of the superradiance likely due to the larger spatial overlap of the exciton wavefunctions between more strongly confined QDs. In contrast to the superradiance, the exciton PL from uncoupled QDs shows constant $g^2(\tau)$ value at all delay times, as expected from random uncoupled emitters (See Fig. S5 in SI). Fig. 2g and 2h compare the time-resolved intensities of the superradiance and PL from uncoupled QDs in the superlattice formed from 9 nm/3C-C8 and 4 nm/3C-C8 QDs, measured at 10 K. 4 nm/3C-C8 QD superlattice shows three-fold faster decay of superradiance (150 ps) than the PL from uncoupled QDs (490 ps) that exhibit the same decay rate of exciton PL from a dilute dispersion of QDs in polymer matrix. This acceleration can be attributed to the increased strength of dipole via coherent coupling of the QDs, similarly to the findings by Blach et al.[17] The acceleration of the decay of superradiance from 9 nm/3C-C8 QDs superlattice is weaker, showing a 1.4-fold acceleration compared to the PL from uncoupled QDs (260 ps vs 370 ps), indicating the weaker coupling than in 4 nm/3C-C8 QDs.

Compared to the superfluorescence previously reported for weakly-confined CsPbBr$_3$ QDs passivated with long ligand,[12, 24] the superradiance from the coupled QDs investigated here exhibits several differences. One is the linear dependence of the superradiance intensity without showing threshold behavior, as the cooperative emission emerges directly from the coherently coupled emitting state. Fig. 3a and 3b show the excitation fluence dependence of the superradiance intensity at 10 K, measured from the superlattice formed from 9 nm/3C-C8 and 4 nm/3C-C8 QDs, exhibiting a linear excitation intensity dependence. Superradiance was observed at the excitation fluence as low as 5 nJ/cm$^2$ (See Fig. S6 in SI). The decay time of the superradiance is also independent of the excitation fluence as shown in Fig. 3c and 3d, in contrast to the superfluorescence that exhibits faster decay with increasing excitation fluence.[12, 15]

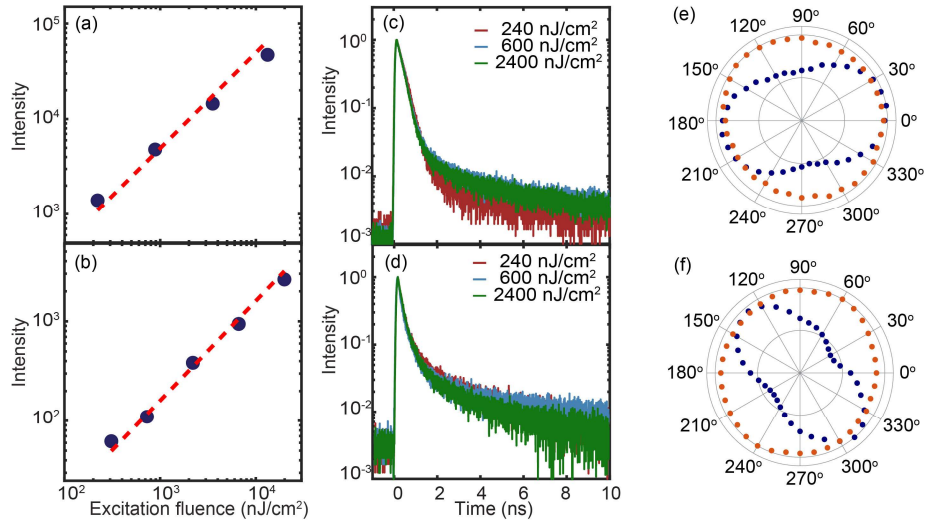

**Fig. 3** (a, b) Excitation fluence dependence of superradiance intensity at 10 K. (c, d) Normalized excitation fluence dependence of superradiance decay at 10 K, (e, f) PL polarization anisotropy of superradiance (blue) and uncoupled exciton PL (red) at 10 K. (a, c, e) Superlattice formed from 9 nm/3C-C8 QDs, and (b, d, f) superlattice formed from 4 nm/3C-C8 QDs.

A unique and unexpected feature observed in the superradiance from this study that has not been previously observed is the polarization anisotropy of the emission. In the absence of intrinsic anisotropy in exciton transition in individual nanocrystals, such as nanoplatelets and nanorods, or geometry of superlattice that introduces anisotropy in the light propagation or interaction with QDs, a superlattice is not expected to exhibit the anisotropy of emission.[25, 26] Cube-shaped CsPbBr$_3$ QDs do not possess a significant anisotropy of the exciton PL due to transition dipoles present along each axis.[27, 28] Therefore, the ordered array of the QDs within the superlattice should not exhibit polarization anisotropy of exciton PL at ambient temperature, as confirmed from a separate experiment (See Fig. S7 in SI). Fig. 3e and 3f compare the polarization anisotropy of the superradiance and uncoupled exciton PL measured at 10 K from superlattices formed from 9 nm/3C-C8 and 4 nm/3C-C8 QDs. Superradiance from both 9 nm/3C-C8 and 4 nm/3C-C8 QD superlattices exhibit a preferred polarization direction, in contrast to the nearly isotropic uncoupled exciton PL. Since both superradiance and uncoupled exciton PL come from the same superlattice, the absence of polarization anisotropy in the uncoupled exciton PL rules out the possibility that the geometry of the superlattice introduces extrinsic anisotropy in the measured PL signal. The polarization angle-dependent superradiance is only in its intensity without any change in spectral shape or peak position (See Fig. S8 in SI). Therefore, this observation is interpreted as arising from anisotropic electronic coupling of the QDs within the superlattice. The preferred polarization direction of the superradiance was independent of the polarization of excitation light at 405 nm that excites above the bandgap of the QDs, and the superradiance intensity did not exhibit dependence on excitation polarization. It is unclear how anisotropic electronic coupling could arise in the superlattice without an immediately identifiable cause that breaks the symmetry. Further studies exploring more detailed structure-polarization property relationship, including the polarization-dependent absorption measurement that is beyond our current instrumental capability, will shed more light on this interesting phenomenon. We conjecture that slight asymmetries in the QD structure and inter-QD interaction may significantly impact the formation of anisotropic coupled dipoles. Nevertheless, the potential to construct polarized superradiant light sources using superlattice fabricated from perovskite QDs could expand the applications of perovskite QDs as the source of quantum photons with controlled characteristics.

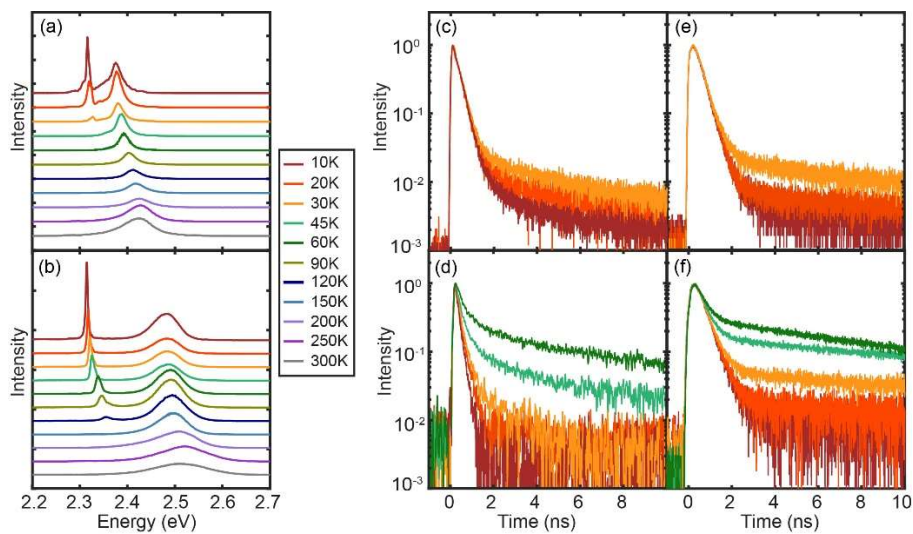

**Fig. 4** (a, b) Temperature-dependent PL spectra of the superlattices formed from (a) 9 nm/3C-C8 and (b) 4 nm/3C-C8 QDs. (c, d) Temperature-dependent decay of superradiance from the superlattices formed from (c) 9 nm/3C-C8 and (d) 4 nm/3C-C8 QDs. (e, f) Temperature-dependent decay of uncoupled exciton PL from (e) 9 nm/3C-C8 and (f) 4 nm/3C-C8 QDs.

The necessary coherence between the QDs to produce superradiance generally decreases with increasing temperature.[29-31] To examine the robustness of the coherence in the superlattice, temperature-dependent PL spectra and PL decay were measured. Fig. 4a and b show the PL spectra in the temperature range of 10-300 K measured from the superlattices of 9 nm/3C-C8 and 4 nm/3C-C8 QDs. The peak attributed to the superradiance begins to appear at 120 K from 4 nm/3C-C8 QD superlattice and its intensity increases with decreasing temperature. In the 9 nm/3C-C8 QD superlattice, the superradiance begins to appear at 30 K, significantly lower than in the 4 nm/3C-C8 QD superlattice. This suggests that the stronger quantum confinement of the QDs that allows the larger exciton wavefunction overlap helps maintain the coherence in the coupled QDs at higher temperatures. In Fig. 4b, the redshift of superradiance with respect to the uncoupled exciton PL increases with decreasing temperature, consistent with the higher degree of coupling of QDs at the lower temperature. The small temperature-dependent shift of the exciton PL peak reflects primarily the variation of the bandgap with the temperature.[30, 32] Temperature-dependent decays of superradiance and uncoupled exciton PL are also compared in Fig. 4c-4f. 4 nm/3C-C8 QD superlattice exhibits significant acceleration of the decay of the superradiance with the decrease of temperature, in contrast to the uncoupled exciton PL that shows substantially weaker temperature dependence. This is consistent with the interpretation of the acceleration of the radiative decay rate resulting from the increase of the dipole strength via coherent coupling of the QDs.[33] The temperature dependence of the peak position and linewidth of both superradiance and uncoupled exciton PL are in SI (See Fig. S9).

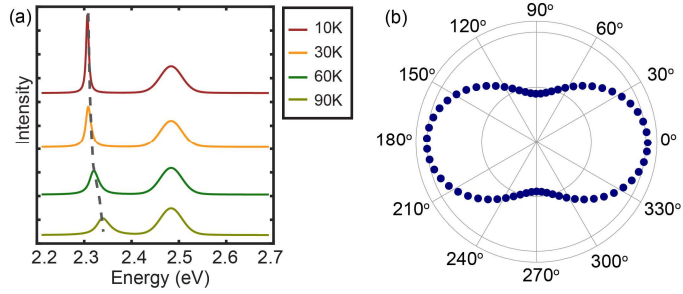

**Fig. 5** (a) Simulated superradiance spectra from the coupled QDs at different temperatures calculated using the best-fit parameters of the tight-binding model to the experimental data. The best-fit parameters are $E_v = 0.015$ eV, $J_x = 0.06$ eV, $J_y = 0.04$ eV, $\varepsilon_0 = 2.5$ eV, $\gamma(T = 0) = 2$ meV. The dashed gray curve shows the measured superradiance peak position. (b) Simulated polarization anisotropy of superradiance with respect to the horizontal x-axis for $d_x/d_y = 3/2$. Other parameters are the same as in panel (a).

The main spectral features of superradiance, their temperature dependence, and the polarization anisotropy observed in this study can be well-formulated using a model described below, which reproduces results that are in qualitative agreement with the experiment. We treat the uncoupled QDs as two-level quantum emitters with a Gaussian distribution of exciton transition energy centered at 2.485 eV with the FWHM linewidth of 52 meV, matching the position and linewidth of the observed broad peak in the PL spectrum between 2.4-2.6 eV. The electronic coupling between QDs is then introduced via the tight-binding Hamiltonian,

$$H = \sum_i \varepsilon_i^c a_{ic}^\dagger a_{ic} + \sum_{ij} J_{ij}^c a_{ic}^\dagger a_{jc} \qquad (1)$$

where $a_{ic}, a_{ic}^\dagger$ are the annihilation and creation operators for the excited state of the *i*th exciton. For convenience, we assume that the ground state of all excitons is the same whereas the excited state energies $\varepsilon_i^c$ differ. Taking into account only nearest-neighbor coupling, we can either numerically diagonalize the Hamiltonian to find eigenenergies for a given number of QDs, or assume a large enough periodic superlattice, so that one can define momentum $\boldsymbol{k} = (k_x, k_y, k_z)$ and go to the limit of the momentum-space Hamiltonian and continuum bands.[34] The energy dispersion in this limit is $\varepsilon_{ck} = \varepsilon_0 - 2J_x cos(k_x a) - 2J_y cos(k_y a) - 2J_z cos(k_z a)$, where the parameters can be found by fitting the numerical results to the observed PL spectra. Here, $a$ is the superlattice period and $J_{x,y,z}$ are the components of the hopping parameter, i.e., essentially the Fourier amplitudes of expanding the interaction Hamiltonian in Eq. (1) in the momentum space. In this model, the narrow PL peak from the coupled QDs is due to optically excited carriers that relax to the bottom of the lowest excited band around $k = 0$ before their recombination. The observed redshift of this peak relative to the uncoupled PL spectrum is determined by the magnitudes of the hopping parameters that vary with temperature. The hopping parameter decreases with increasing temperature due to vibrational (phonon) excitations, which are activated with probability $e^{-E_v/(k_B T)}$, where $E_v$ is the characteristic vibrational energy and $k_B$ is the Boltzmann constant. Therefore, the hopping parameter scales with temperature as $J_{x,y}(T) = J_{x,y}(0)\left(1 - e^{-\frac{E_v}{k_B T}}\right)$. Within this model, there is no qualitative difference between the PL from 2D and 3D superlattices, other than the rescaling of the numerical values of the hopping parameters. Therefore, to save computation time, we used a 2D model with two components $J_x$ and $J_y$ having different values to account for the observed anisotropy as explained below. At low temperatures, the width of the PL peak from the coupled QD state is determined by the homogeneous linewidth $\gamma(T)$ at temperature $T$. Its value decreases linearly with decreasing temperature due to reduced electron-phonon scattering.[35] The resulting redshift and narrowing of the superradiance calculated for the 4 nm/3C-C8 QD superlattice are shown in Fig. 5a, which are in good agreement with the data in Fig. 4b. The difference between 4 nm/3C-C8 QD and 9 nm/3C-C8 QD superlattices can be explained by lower inter-band transition energy and smaller hopping parameters for the larger QDs and difference in phonon density of states. The anisotropy of superradiance cannot be explained by the shape of the superlattice or the intrinsic anisotropy of the individual QDs as mentioned earlier. Therefore, we assumed that it results from the anisotropy in the inter-QD coupling, i.e., unequal values of the hopping parameters $J_x$ and $J_y$ and resulting optical dipole matrix elements $d_{x,y} \propto J_{x,y}$. The observed polarization anisotropy of superradiance shown in Fig. 3f is best reproduced with $J_x/J_y \approx 3/2$ as shown in Fig. 5b, ignoring the actual direction of the anisotropy axis in the superlattice.

In conclusion, we report the observation of polarized superradiance from the electronically coupled CsPbBr$_3$ perovskite QDs in superlattice. Unlike in superfluorescence emerging from the buildup of coherence from the incoherently excited emitters, the superradiance from the coupled QDs is governed by inter-QD electronic coupling that can be tuned by the quantum confinement and ligand engineering of the QDs forming the superlattice. We observed that the combination of strong quantum confinement and the use of the shorter ligand not only enhances the inter-QD coupling but also introduces anisotropy in the coupling, enabling the polarized cooperative photon emission. These results demonstrate the potential of the ordered assembly of perovskite QDs with controllable electronic coupling as the source of the polarized superradiant light.

# Methods

## 1. Synthesis of CsPbBr$_3$ quantum dots (QDs) and superlattice fabrication

**Materials.** Cesium carbonate (99.994%, Alfa Aesar), lead bromide (98%, Alfa Aesar), zinc bromide (99%, BeanTown Chemical), hydrobromic acid (48%, VWR), 1-octadecene (90% technical grade, Sigma-Aldrich), oleylamine (≥98%, Sigma Aldrich), oleic acid (90% technical grade, Sigma-Aldrich), 1,3-dibromopropane (98% TCI), *N*,*N*-dimethyloctylamine (97% Sigma-Aldrich), *N*,*N*-dimethyldodecylamine (96% TCI), were used as received without further purification.

**Synthesis of 9 nm CsPbBr$_3$ QDs passivated with oleylammonium bromide (OLAB).** The synthesis procedure was adopted from a report by Yakunin et al.[36] Cesium oleate was prepared by combining Cs$_2$CO$_3$ (250 mg), oleic acid (OA, 0.8 g) and 1-octadecene (ODE, 7 g) in a 50 ml three-neck round-bottomed flask under a nitrogen atmosphere. The flask was evacuated at room temperature for 10 minutes before being heated to 150 °C for an additional 10 minutes. The flask was refilled with nitrogen and held at 120 °C for further use. In another 50 mL round-bottomed flask, PbBr$_2$ (140 mg), OA (1 mL), oleylamine (OAm, 1 mL), and ODE (10 mL) were combined under nitrogen and stirred at room temperature for 10 minutes. The flask was then evacuated for 20 minutes before being refilled with nitrogen. It was heated to 120 °C until the precursors fully dissolved and heated further to 220 °C. To initiate the reaction, 0.8 mL of Cs-oleate was swiftly injected into the flask. The reaction was allowed to proceed for 5 seconds before quenching with an ice bath. To purify the resulting QDs, the crude solution was centrifuged to precipitate the QDs and the recovered QDs were redisperse in hexane. The QDs were precipitated again by adding methyl acetate to the hexane solution, and the recovered QDs were resuspended in hexane for further use.

**Synthesis of 4 nm CsPbBr$_3$ QDs passivated with OLAB.** The synthesis procedure was adopted from a report by Dong et al.[37] Cesium oleate was prepared by combining Cs$_2$CO$_3$ (600 mg), OA (2.4 mL) and ODE (6.4 mL) in a 50 ml three-neck round-bottomed flask under a nitrogen atmosphere. The flask was evacuated at room temperature for 15 minutes before being heated to 120 °C for an additional 15 minutes. The flask was refilled with nitrogen and held at 120 °C for further use. In a separate 100 mL round-bottomed flask, PbBr$_2$ (350 mg), ZnBr$_2$ (700 mg), OA (7 mL), OAm (7 mL) and ODE (20 mL) were mixed under nitrogen and stirred at room temperature for 10 minutes. The flask was then evacuated for 20 minutes before being refilled with nitrogen. The flask was heated to 150 °C for 20 minutes and cooled to 100 °C. To initiate the reaction, 3 mL of Cs-oleate was swiftly injected into the flask. The reaction was allowed to proceed for an hour before being quenched with an ice bath. To purify the resulting QDs, acetone was added to the crude solution to precipitate the QDs. Supernatant was separated and discarded via centrifugation, and the precipitate was resuspended in hexane for further use.

**Synthesis of 3C-C8 Ligands.** The 3C-C8 ligand was synthesized using the method adapted from a previous report.[38] A bidentate ligand was chosen because it was effective in providing the stability of the QDs and superlattice when short-chain ligand is used.[38,39] Tertiary amine (8 mmol) and 1,3-dibromopropane (0.404 g, 2 mmol) were dissolved in 5 mL acetonitrile in a 25 mL round-bottomed flask. A condenser was connected to the flask and the flask was purged with nitrogen. The mixture was refluxed overnight under a nitrogen atmosphere, then cooled to room temperature. The crude product was precipitated by centrifugation after adding diethyl ether. Additional diethyl ether was added to the precipitate and sonicated for 15 minutes before isolating the product by centrifugation. The resulting white powder was dried in vacuo overnight. Recrystallization with toluene was performed to purify the ligands before use. The identity and purity of the product were confirmed by nuclear magnetic resonance (NMR) spectroscopy. The NMR spectrum is shown in Fig. S1 in SI.

**Ligand exchange of OLAB-passivated $CsPbBr_3$ QDs with 3C-C8.** All $CsPbBr_3$ QDs passivated with 3C-C8 ligand were prepared via ligand exchange of the OLAB-passivated $CsPbBr_3$ QDs. 250 μL of concentrated OLAB-passivated $CsPbBr_3$ QD solution (≥0.25 mM) was transferred to a centrifuge tube. Methyl acetate was then added to the solution to bring the total volume to 2 mL, partially stripping the native ligand from the QDs. The QDs were isolated from the solution via centrifugation. The precipitated QDs were redispersed in 100 μL of a toluene solution containing 3C-C8. The solution was vigorously mixed for 5 minutes. Methyl acetate was again added to bring the total volume to 2 mL, followed by centrifugation to obtain the pellet. The pellet was redispersed in toluene until it dissolved completely. The solution was centrifuged at 6000 rpm for 3 minutes, and the supernatant was collected. The exchange process was repeated three times to complete the exchange.

**Fabrication of superlattice of $CsPbBr_3$ QDs.** The superlattices of $CsPbBr_3$ QDs were fabricated via slow solvent evaporation following a previously reported procedure with modifications.[12] A silicon wafer used as a substrate (1cm ×1cm) was cleaned with sonication in a mixture of acetone and isopropyl alcohol. To remove the residual solvent and moisture, the substrate was dried with nitrogen gas followed by further drying in an oven. To form the superlattice, 40 μL of the diluted QD solution in toluene was dropped onto the silicon wafer and sealed in a PTFE well covered with a glass slide. The well was then transferred into a desiccator purged with nitrogen, where the self-assembly of the QDs was allowed to continue for 36 hours to form μm-sized cuboidal superlattices. The superlattices formed on the silicon wafer were used for all the optical measurements. The superlattices grown on TEM grids employing the same method were used for imaging with a transmission electron microscope.

**Fabrication of dilute dispersion of QDs in polystyrene matrix.** To fabricate the isolated QDs dispersed in polystyrene matrix, a dilute QD solution was prepared using 6% polystyrene (MW 92,000) solution in toluene. The resulting optical density in 1 cm cuvette was ~1 at the exciton absorption peak. The diluted solution was directly drop-cast on a 1cm×1cm silicon wafer. The solvent was then removed in a vacuum oven, forming a uniform thin film of QDs dispersed in the polymer matrix.

**Transmission electron microscopy of QDs.** QD specimens were prepared by dropping 2.0 μL of concentrated QD solution onto carbon-supported copper grids (Ted Pella, 01840-F) pretreated with oxygen plasma (SolarusII, Gatan Inc., 20 W for 15 s) and drying under argon atmosphere. TEM images were acquired using a JEM 2100Plus (JEOL) electron microscope equipped with a scintillator-based camera (Gatan, OneView) and a specimen double-tilting holder (JEOL, EM-31630), operating at an acceleration voltage of 200 kV.

## 2. Spectroscopic characterization

**Solution-phase absorption and PL spectra.** Solution-phase absorption spectra of the QDs were obtained with a CCD spectrometer (Ocean Optics, USB2000) equipped with a deuterium light source (UV-VIS ISS, Ocean Optics). Steady-state PL spectra of the QD solutions were obtained with a CCD spectrometer (Ocean Optics, USB2000) and a 365 nm UV light emitting diode as an excitation source.

**Steady-state PL spectra of the superlattice and dilute dispersion of QDs.** Steady-state PL spectra of superlattice and dilute dispersion of QDs in polymer matrix were measured using a home-built confocal microscope equipped with a dual grating spectrograph (Princeton Instruments, SP-2150i) and a CCD camera (Princeton Instruments, PIXIS 100). (See Fig. S12) A sample prepared on a silicon substrate was placed inside an open-cycle helium optical cryostat (JanisST-500), where the temperature was controlled between 10-300 K with a temperature controller (Lakeshore, Temperature Controller 332). A 405 nm diode laser (PicoQuant, P-C-405) was used as an excitation source.

**Time-correlated single photon counting measurements.** Time-resolved PL intensity of the superlattice and dilute dispersion of QDs in polymer matrix were measured using a time-correlated single photon counting setup that consists of a time correlation module (PicoQuant, PicoHarp 300) and an avalanche photodiode (PicoQuant, MPD, PDM series). A 405 nm pulsed diode laser (PicoQuant, P-C-405) producing pulses of 45 ps width was used as the excitation source at a repetition rate of 5 MHz.

**Second-order photon correlation measurement.** Hanbury Brown-Twiss (HBT) interferometer setup was constructed for the measurement of second-order photon correlation. Two identical avalanche photodiodes (PicoQuant, MPD, PDM series) were used to detect the photons on each arm of the interferometer. A 405 nm pulsed diode laser (PicoQuant, P-C-405) producing pulses of 45 ps width was used as the excitation source at a repetition rate of 80 MHz.

**PL and excitation polarized-dependent measurements.** For the measurement of PL polarization anisotropy, a linear polarizer (Thorlabs, LPVISA100) was placed in front of the avalanche photodiodes and spectrometer. For the measurement of the excitation polarization-dependent PL spectra, the combination of a polarizing beamsplitter cube (Thorlabs, PBS251) and a half waveplate (Thorlabs, WPH10M-405) was used to vary the polarization of 405 nm excitation light.

## Data availability
All the data supporting this article are included in the main text and supplementary Information. Raw data can be obtained from the corresponding author on request.

## Acknowledgments

This work was supported by the National Science Foundation (CHE-2003961 to DHS), the Air Force Office for Scientific Research Grant (FA9550-21-1-0272 to AB) and Texas A&M University (STRP).

## Author contributions
L.L carried out sample growth, experiments and data analysis. X.T and M.K conducted the early-stage test experiments. J.P, and C.W.W. carried out the material synthesis and obtained TEM images of the solution-phase samples. M.P and J.C carried out TEM imaging of the superlattices. A.S. and A.B. developed the model and performed simulations. L.L, A.V.S, A.B and D.H.S wrote the manuscript.

## Competing interest
The authors declare no competing interests.

## Additional Information
**Correspondence and requests for materials** should be addressed to Dong Hee Son.

# Supplementary Information
# Polarized Superradiance from CsPbBr$_3$ Quantum Dot Superlattice with Controlled Inter-dot Electronic Coupling

*Lanyin Luo[1,2], Xueting Tang[3], Junhee Park[3], Chih-Wei Wang[3], Mansoo Park[4], Mohit Khurana[1,2], Ashutosh Singh[1], Jinwoo Cheon[4], Alexey Belyanin[1], Alexei V. Sokolov[1,2] and Dong Hee Son[1,3,4]*

[1]Department of Physics and Astronomy, Texas A&M University, College Station, TX 77843, USA
[2]Institute of Quantum Science and Engineering, Texas A&M University, College Station, TX 77843, USA
[3]Department of Chemistry, Texas A&M University, College Station, Texas 77843, United States
[4]Center for Nanomedicine, Institute for Basic Science and Graduate Program of Nano Biomedical Engineering, Advanced Science Institute, Yonsei University, Seoul 03722, Republic of Korea

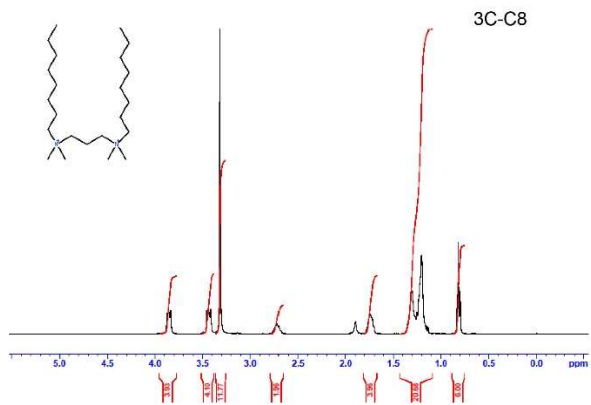

**Figure. S1.** $^1$H NMR of 3C-C8 ligand synthesized.
**3C-C8**: $^1$H NMR (400MHz, CDCl$_3$): 3.87-3.83 (t, J = 7.8 Hz, 4H), 3.80-3.50 (m, 4H), 3.32 (s, 12H), 2.78-2.66 (m, 2H), 1.81-1.67 (m, 4H), 1.73-0.99 (m, 20H), 0.83-0.79 (t, J = 5.5 Hz, 6H)

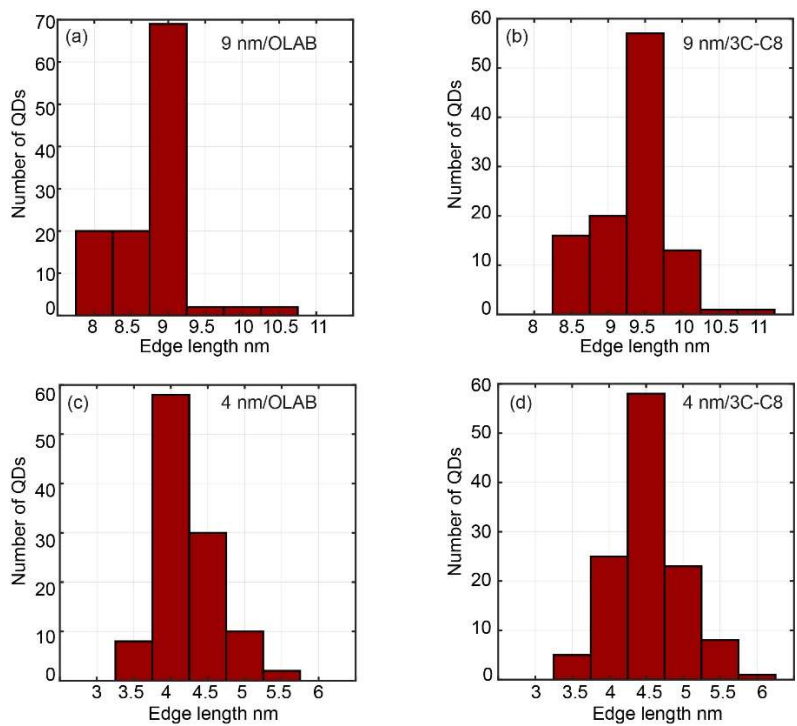

**Figure. S2.** Histogram of the NC sizes obtained from the TEM images in Fig. 1. (a) 9 nm QDs passivated with OLAB, (b) 9 nm QDs passivated with 3C-C8, (c) 4 nm QDs passivated with OLAB, (d) 4 nm QDs passivated with 3C-C8.

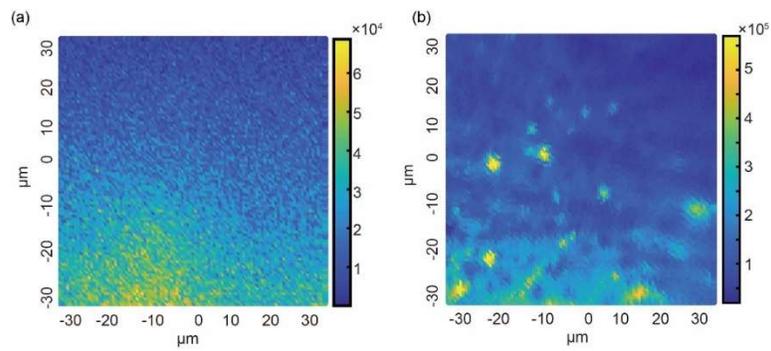

**Figure. S3.** Representative confocal fluorescence image of (a) dilute dispersion of CsPbBr$_3$ QDs in polystyrene matrix and (b) CsPbBr$_3$ QD superlattices obtained at 10 K.

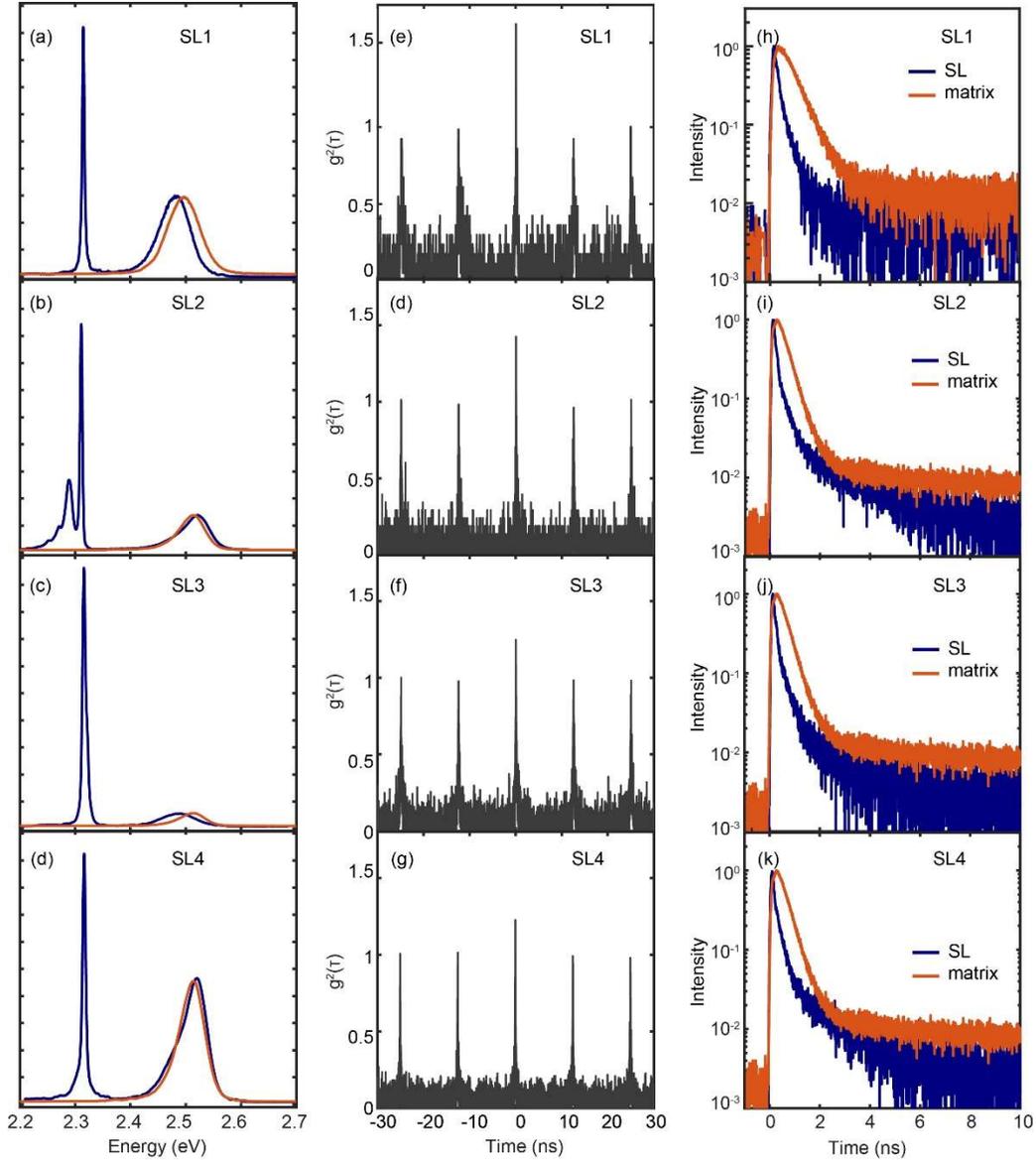

**Figure. S4.** Variations of superradiance among four different superlattices (SL1-SL4) formed from 4 nm/C3-C8 CsPbBr$_3$ QDs. (a-d) PL spectra of superlattices (blue) and dilute dispersion (red) at 10 K. (e-g) Second-order photon correlation, g$^2$(τ), of superradiance from superlattice. (h-k) Comparison of the normalized time-dependent PL intensities from superradiance and PL from uncoupled QDs. All data are obtained at 10 K under 405 nm pulsed excitation at the fluence of 240 nJ/cm$^2$. FWHM of superradiance varies in 3-5 meV range. The redshift of superradiance from the uncoupled exciton PL varies in 180-220 meV range. Significant photon bunching with $g^2(0) > 1.2$, and accelerated lifetime were detected, indicating cooperative emission from 4 nm/3C-C8 superlattice.

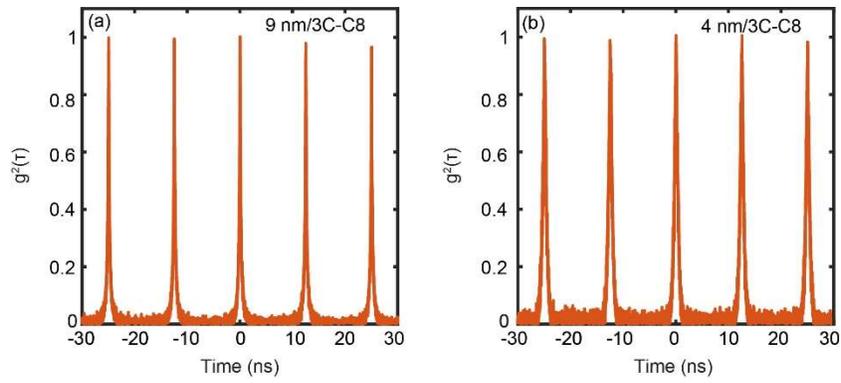

**Figure. S5**. Second order photon correlation, $g^2(\tau)$ of uncoupled exciton PL from the superlattice formed from (a) 9 nm/3C-C8 QDs and (b) 4 nm/3C-C8 QDs measured at 10 K.

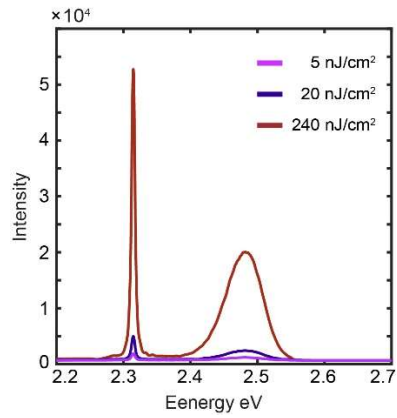

**Figure. S6.** Excitation fluence-dependent PL spectra of the superlattice formed from 4 nm/3C-C8 QDs at 10K. Superradiance does not show any changes in the spectra shape and the peak position.

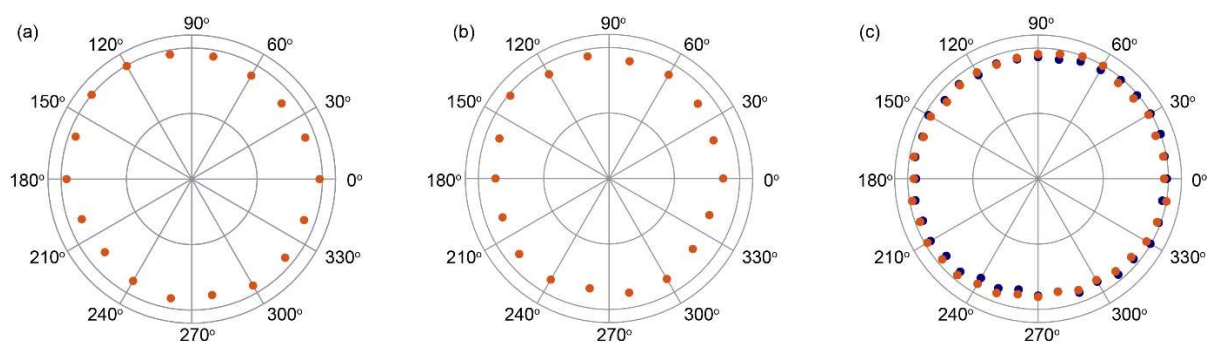

**Figure. S7.** (a, b) Polarization anisotropy of uncoupled exciton PL from the superlattices formed from (a) 9 nm/3C-C8 QDs and (b) 4 nm/3C-C8 QDs under 405 nm excitation in ambient temperature. (c) Excitation polarization-dependence intensity of superradiance (blue) and uncoupled exciton PL (red) from the superlattice formed from 4 nm/3C-C8 QDs at 10 K.

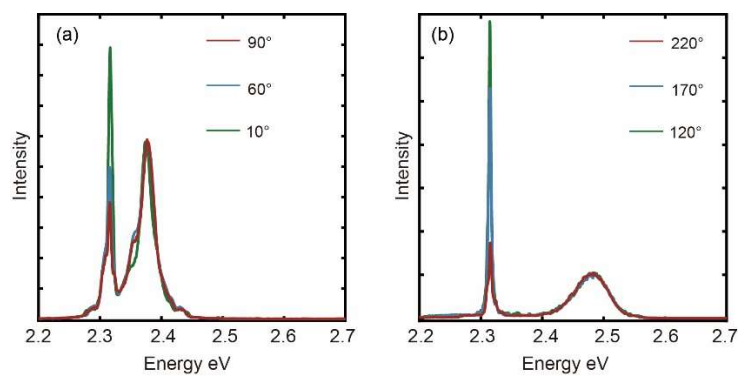

**Figure. S8.** Polarization-dependent PL spectra of superlattice formed from (a) 9 nm/C3-C8 QDs and (b) 4 nm/C3-C8 QDs at 10 K.

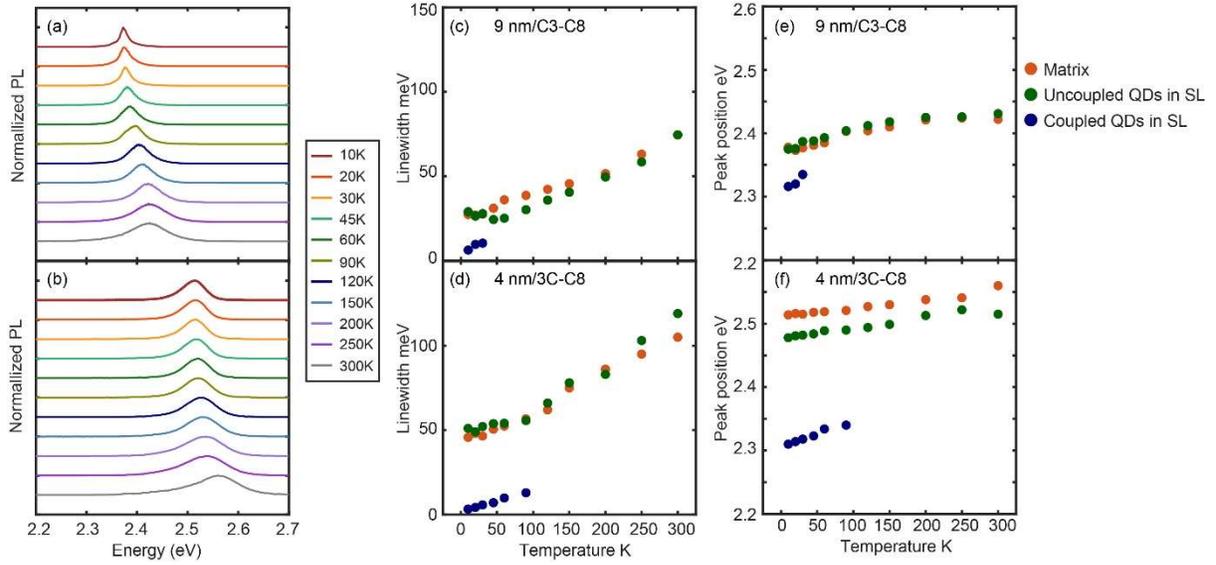

**Figure. S9.** (a, b) Temperature-dependent PL spectra of the dilute dispersion of $CsPbBr_3$ QDs in polystyrene matrix. (a) 9 nm/3C-C8, (b) 4 nm/3C-C8. (c, d) FWHM linewidth of QDs in superlattice and dilute dispersion extracted from Fig. 4 in the main text and Fig. S9(a-b). (c) Linewidth of 9 nm/3C-C8. (d) Linewidth of 4 nm/3C-C8. (e, f) Peak position of QDs in superlattice and dilute dispersion extracted from Fig. 4 in the main text and Fig. S9(a-b). (e) Peak position of 9 nm/3C-C8. (f) Peak position of 4 nm/3C-C8.

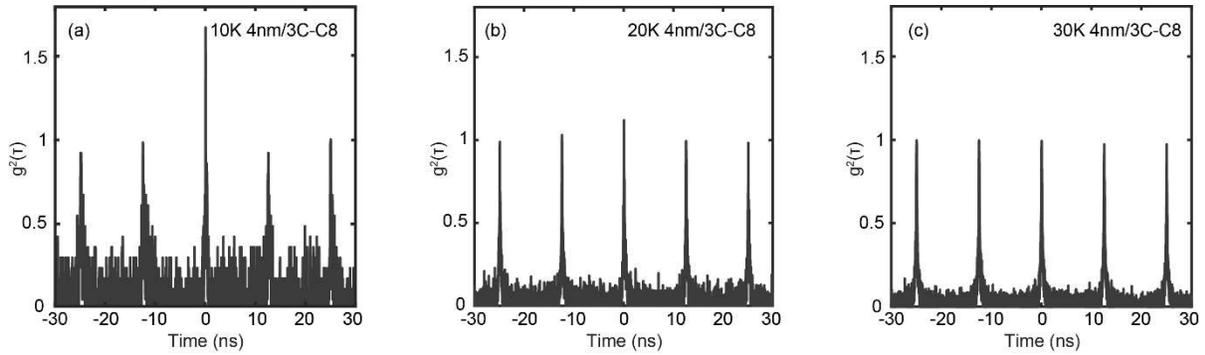

**Figure. S10.** Temperature-dependent second-order photon correlation $g^2(\tau)$ of superradiance from the superlattice formed from 4nm/3C-C8 QDs at (a) 10 K, (b) 20 K, (c) 30 K.

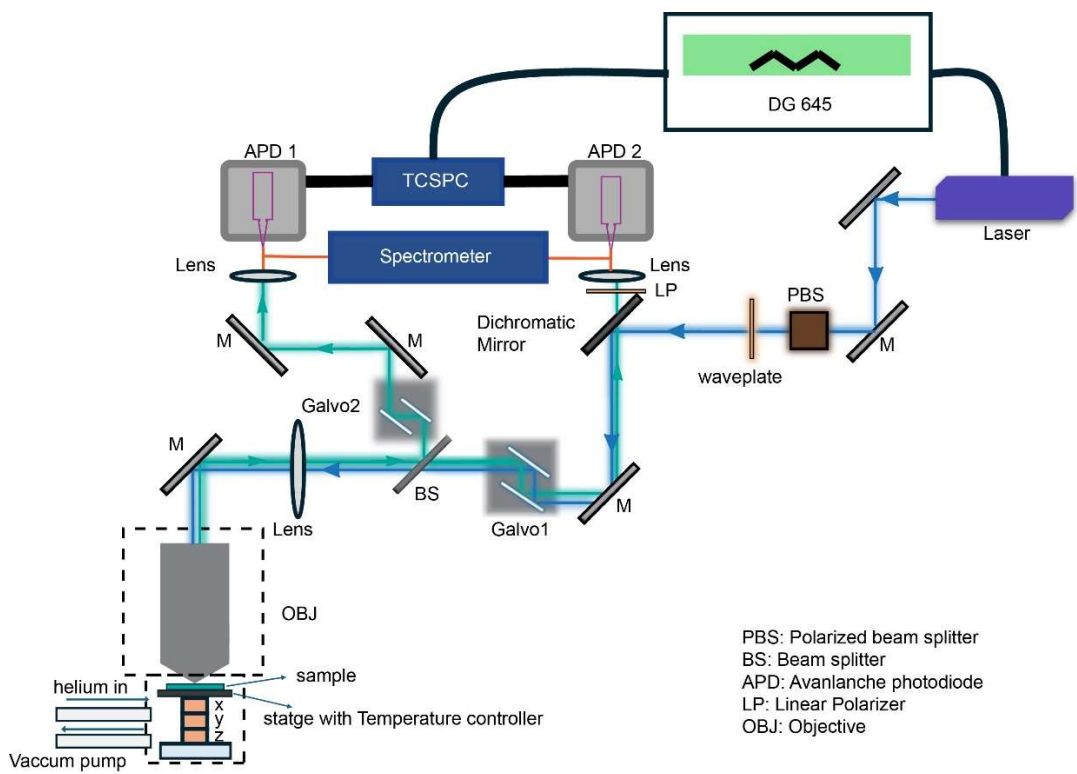

**Figure. S11.** Schematic diagram of the confocal microscope and spectroscopic measurement setup.